\documentstyle[12pt]{article}

\newcommand{\be}{\begin{equation}}
\newcommand{\ee}{\end{equation}}
\newcommand{\vp}{\varphi}
\newcommand{\ra}{\rightarrow}
\newcommand{\lbd}{\lambda}
\newcommand{\al}{\alpha}
\newcommand{\bt}{\beta}

\begin{document}

\begin{center}
{\Large \bf Self-Similar Factor Approximants} \\ [5mm]
{\large S. Gluzman$^1$, V.I. Yukalov$^2$ and D. Sornette$^{1,3,4}$} \\ [3mm]

{\it 
$^1$ Institute of Geophysics and Planetary Physics \\
University of California, Los Angeles, California 90095 \\ [2mm]

$^2$ Bogolubov Laboratory of Theoretical Physics \\
Joint Institute for Nuclear Research, Dubna 141980, Russia\\ [2mm]

$^3$ Department of Earth and Space Science\\
University of California, Los Angeles, California 90095 \\ [2mm]

$^4$ Laboratoire de Physique de la Mati\`ere Condens\'ee \\
CNRS UMR6622  and Universit\'e des Sciences \\
Parc Valrose, 06108 Nice Cedex 2, France}

\end{center}

\vskip 2cm

\begin{abstract}

The problem of reconstructing functions from their asymptotic expansions 
in powers of a small variable is addressed by deriving a novel type of 
approximants. The derivation is based on the self-similar approximation 
theory, which presents the passage from one approximant to another as 
the motion realized by a dynamical system with the property of group 
self-similarity. The derived approximants, because of their form, are 
named the self-similar factor approximants. These complement the 
obtained earlier self-similar exponential approximants and self-similar 
root approximants. The specific feature of the self-similar factor 
approximants is that their control functions, providing convergence of 
the computational algorithm, are completely defined from the 
accuracy-through-order conditions. These approximants contain the Pad\'e
approximants as a particular case, and in some limit they can be 
reduced to the self-similar exponential approximants previously
introduced by two of us. It is proved that 
the self-similar factor approximants are able to reproduce exactly a wide 
class of functions which include a variety of transcendental functions. 
For other functions, not pertaining to this exactly reproducible class, 
the factor approximants provide very accurate approximations, whose 
accuracy surpasses significantly that of the most accurate Pad\'e 
approximants. This is illustrated by a number of examples showing the
generality and accuracy of the factor approximants even when conventional
techniques meet serious difficulties.

\end{abstract}

\newpage

\section{Introduction}

The problem of reconstructing functions from their perturbative asymptotic
expansions in powers of a parameter or a variable is so frequently met in
physics and in applied sciences that there is no necessity to explain its
importance. The best known methods for such a reconstruction are the Pad\'e
approximation and Borel summation, including their variants and combinations
[1,2]. These techniques usually require that a large number of terms of an
asymptotic expansion be available. The Borel summation demands, in addition,
that the high-order expansion coefficients be given and the analytic
properties of the sought function on the complex plane be prescribed.
However, the overwhelming majority of realistic physical problems are too
complicated and perturbation theory is only able to derive a few first
terms. And the luxury of knowing in advance the analytic properties of an
unknown function, together with its high-order expansion coefficients, as is
required for the Borel summation, is practically never available. Because of
the latter, Pad\'e approximants are more often employed in applications,
although their usage also confronts with several difficulties, among which
the most notorious are the appearance of spurious poles and the poor
recovery of non-integer critical exponents.

An alternative approach to the problem of reconstructing functions has been
developed, whose basic ideas are as follows. First of all, to improve the
convergence property of a perturbative sequence, it is necessary to
introduce {\it control functions} defined by an optimization procedure
[3--5]. This idea makes the foundation of the {\it optimized perturbation
theory} that is now widely employed for various applications [3--16]. The
second pivotal idea is to consider the successive passage from one
approximation to the next one as a dynamical evolution on the manifold of
approximants, which is formalized by the notion of {\it group self-similarity%
} [17--22]. And the third principal point is the introduction of control
functions in the course of rearranging perturbative asymptotic expansions by
means of {\it algebraic transforms} [23--27]. Because of their specific
scaling properties typical of fractals [28,29], the algebraic transforms can
also be called {\it fractal transforms} [30]. By using this technique, two
types of approximants have been obtained, {\it self-similar exponentials}
and {\it self-similar roots} [24--27].

In the present paper, we suggest a different approximation scheme resulting
in what may be named {\it self-similar factor approximants}. These new
approximants possess an important principal property distinguishing them
from the self-similar approximants mentioned above: the control parameters,
entering the self-similar factors, can be completely defined from a given
asymptotic expansion by the so-called ``accuracy-through-order'' matching
method. This is in contrast with the self-similar exponentials whose
controls, designed to improve convergence, are defined from additional
optimization conditions. This method of accuracy-through-order is also
different from the determination of the control parameters of the
self-similar roots which are determined by matching two asymptotic
expansions valid in the neighborhood of two different asymptotic points.
Being based on the sole initial asymptotic expansion, the self-similar
factors have the advantage of simplicity which makes their usage quite
attractive. Furthermore, these approximants allow one to reconstruct {\it %
exactly} a wide class of functions. And when they do not yield exact
answers, they provide very accurate approximations, essentially more
accurate than given by Pad\'e approximants.

We, first, give in the next section the mathematical foundation of the
self-similar factor approximants. Sections 3 and 4 are devoted to
examples chosen both for their illustrative properties, the difficulties
they pose to the more convential Pad\'e technique and the relevance to
several physical problems (polymers, state- and velocity-dependent solid
friction dynamics, critical phenomena in field theory and in Ising models).

\section{Mathematical Foundation}

\subsection{Derivation of factor approximants}

Assume that we are solving a complicated problem, aiming at finding a real
function $f(x)$ of a real variable $x$. Because of the complexity of the
problem, the only thing we are able to do is to invoke a kind of
perturbation theory for obtaining approximate expressions $f_k(x)$ of order $%
k=0,1,2,\ldots$, valid in the asymptotic vicinity of $x=0$. Usually, the
asymptotic approximants $f_k(x)$ can be presented as a power series of $x$
and written in the form 
\be
\label{1} 
f_k(x) = f_0 \; \vp_k(x) \; , \qquad \vp_k(x) =\sum_{n=0}^k a_n \;
x^n \; , 
\ee
where $\vp_k(x)$ is a dimensionless function with $a_0=1$. Writing down the
sought function as an asymptotic series 
\be
\label{2} 
f(x) \simeq f_0 \sum_{n=0}^k a_n \; x^n + \ldots \; , 
\ee
where $x\ra 0$, gives little consolation, since in real problems $x$ is
rarely asymptotically small, but usually it is finite and may be even very
large. How could we reconstruct the function $f(x)$ for finite values of $x$
from the only knowledge of its asymptotic expansion?

An answer to this question can be provided by the self-similar approximation
theory [17--22], with control parameters introduced by means of the fractal
transform [23--27,30], defined as 
\be
\label{3} 
F_k(x,s) \equiv x^s\; f_k(x) \; . 
\ee
This leads to the self-similar exponential and self-similar root
approximants. Now, we shall follow a slightly different procedure, which is
actually motivated by the very idea of group self-similarity underlying the
construction of self-similar approximants.

For a more efficient use of the group self-similarity, we propose to present
an initially given asymptotic expansion in the most symmetric way. To this
end, we introduce the {\it factor functions} 
\be
\label{4} 
\vp_{kp}(x) \equiv 1 + b_{kp}\; x \qquad (p=1,2,\ldots, \; k\geq
1) \; . 
\ee
Let us consider the finite series (1) as a polynomial over the field of
complex numbers. Then, by the fundamental theorem of algebra [31], such a
polynomial can be split in a unique way into a product of the irreducible
factors (4), so that 
\be
\label{5} 
\vp_k(x) = \prod_{p=1}^k \vp_{kp}(x) \qquad (k=1,2,\ldots) \; . 
\ee
The representation (5) possesses the scaling property, for which if $\vp_{kp}
\ra \lbd\vp_{kp}$, then $\vp_k\ra\lbd^k\vp_k$. Such a scaling property is
the simplest rudimental example of functional self-similarity. In this way,
the sum (2) can be rewritten as a product \be
\label{6} 
f(x) \simeq f_0 \prod_{p=1}^k \left [ \vp_{kp}(x) + \ldots \right ] \; . 
\ee
Now, instead of accomplishing the self-similar summation for the whole
right-hand side of Eq. (2), we may perform it for each factor in the product
(6). Thus, we define the fractal transform 
\be
\label{7} 
\Phi_{kp}(x,s) \equiv x^s \; \vp_{kp}(x) \; , 
\ee
construct an approximate cascade whose trajectory is bijective to the
sequence $\{\Phi_{kp}(x,s)\}$, embed the cascade into an approximation flow,
integrate the flow evolution equation, and realize the inverse fractal
transform. All this machinery, with all details, has been expounded in
previous papers [17--27], and we therefore do not repeat it here. As a
result, each factor (4) can be shown to be renormalized into 
\be
\label{8} 
\vp_{kp}^*(x) = \left ( 1 + A_{kp}\; x\right )^{n_{kp}} \; , 
\ee
where $A_{kp}$ and $n_{kp}$ are control parameters, or simply controls. And
a $k$-order approximation for the sought function $f(x)$ is given by the 
{\it self-similar factor approximant} 
\be
\label{9} 
f_k^*(x) = f_0 \prod_{p=1}^k \vp_{kp}^*(x) \; . 
\ee
The controls $A_{kp}$ and $n_{kp}$ are determined by expanding the
approximant (9) in powers of $x$ and comparing this expansion with the
series (2). For short, this can be called a re-expansion procedure, which
sometimes is also named the accuracy-through-order relationship. The
equations defining the amplitudes $A_{kp}$ and exponents $n_{kp}$ can be
cast in the form 
\be
\label{10}
\sum_{p=1}^k n_{kp} \; A_{kp}^n = (-1)^{n+1} \; n\; b_n \; , 
\ee
where 
\be
\label{11}
b_n \equiv \frac{1}{n!}\; \lim_{x\ra 0}\; \frac{d^n}{dx^n}\; \ln \left (
\sum_{m=0}^\infty a_m\; x^m \right ) \; . 
\ee
As is easy to check, Eqs. (10) and (11) follow from equating the asymptotic
expansions for the logarithms of the factor approximant (9) and of series 
(2). For each given $k$, there are $2k$ unknowns in the left-hand side of 
Eq. (10). Hence, in the series (2), one should have $2k$ nontrivial 
terms, which makes $n=1,2,\ldots,2k$. The series of odd orders $2k+1$
can also be processed, for which one needs to consider $f(x)-f_0$, instead
of $f(x)$. The factor approximants, based on even and odd numbers of terms
of a series with alternating signs, often bracket the sought function from
below and above. This bracketing is analogous to that occurring for
self-similar exponential approximants based on even or odd numbers of
asymptotic terms [26,32].

The control parameters may be complex-valued, since, for obtaining the
factorized form (5), the sum (1) was treated as a polynomial over the field
of complex numbers. But, since the considered function is real, all
complex-valued factors should arise in complex conjugate pairs, so that
their product be always real.

\subsection{Exactly reproducible class of functions}

The structure of the self-similar factor approximants (9) suggests that
there exists a whole class of functions that are exactly reproducible by
means of these approximants. This class is defined as follows. Let $P_n(x)$
be an irreducible polynomial in a real variable $x$ of degree $n$ over the
field of real numbers and let $\al_i$ and $\bt_j$ be complex numbers.
Compose the real-valued products of powers of such irreducible polynomials
as $\prod_i P_{M_i}^{\al_i}(x)$ and $\prod_j Q_{N_j}^{\bt_j}(x)$, where 
$\sum_i M_i=M$ and $\sum_j N_j=N$. This implies that complex powers, if any,
always come in complex conjugate pairs. Let these products have no common
divisors, such that the ratio 
\be
\label{12} 
f_{MN}^{\al\bt}(x) \equiv 
\frac{\prod_i P_{M_i}^{\al_i}(x)}{\prod_j Q_{N_j}^{\bt_j}(x)} 
\ee
be irreducible. Denote by ${\cal R}$ a class of functions, which is composed
of all products of the forms (12) that play the role of the prime
representatives for this class.

\vskip 3mm

{\it Theorem}. A function $f(x)$ can be exactly reproduced by the
self-similar factor approximants (9) if and only if this function belongs to
the class ${\cal R}$, with the prime representatives (12) being exactly
reproducible by $f_k^*(x)$ in all orders $k\geq M+N$.

\vskip 2mm

{\it Proof}. To prove the proposition for the whole class ${\cal R}$, it is
necessary and sufficient to prove it for the prime representatives (12).
According to the fundamental theorem of algebra [31], each polynomial in one
real variable, over the field of real numbers, can be split into factors of
the first and second degree, over the field of real numbers, and into
factors of the first degree, over the field of complex numbers. This allows
us to split each polynomial entering the prime representative (12) into the
product of first-degree factors with complex coefficients. Since, by
definition, the ratio (12) is irreducible, it can be presented as a product
of $M+N$ first-degree factors. After this, it acquires the form identical to
the factor approximant (9) of the order $k=M+N$, taking into account that
the exponents $n_{kp}$ in (\ref{8}) can be negative. Hence the latter
pertains to the class ${\cal R}$. And if a function is exactly reproducible 
by an approximant (9), this function must be from the class ${\cal R}$.

By construction, all parameters (coefficients and powers) of the approximant
(9) are defined by equating its asymptotic, in $x\ra 0$, expansion with that
of the function (12). An asymptotic expansion in the sense of Poincar\'e is
uniquely defined by the function itself. Two functions, having an identical
dependence on the variable and coinciding asymptotic expansions, coincide.
That is, a function from the class ${\cal R}$ is exactly reproducible by a
factor approximant (9). Finally, if a function $f(x)$ exactly equals an
approximant $f_k^*(x)$ of the order $k=p$, then the higher approximants,
with $k>p$, derived from the same asymptotic expansion of the same function $%
f(x)$, will coincide with each other.

\vskip 2mm

{\it Remarks}. The class ${\cal R}$ of the functions, exactly reproducible
by means of the self-similar factor approximants, is significantly wider
than the class of rational functions that can be exactly reproduced by
Pad\'e approximants. In addition to rational functions, the class ${\cal R}$
also includes transcendental functions. Because of this, the self-similar
factor approximants should provide a better accuracy for a wider class of
functions, as compared to Pad\'e approximants. In what follows, we shall
illustrate this by a variety of examples.

For a given real-valued asymptotic series (2) of a real function $f(x)$, the
factor approximants (9), by construction, are real in the asymptotic region
of $x\ra 0$. However, they may become complex for finite $x$. If an
approximant $f_k^*(x)$ becomes complex for $x>x_k$, this implies that the
region of validity of $f_k^*(x)$ is restricted by the interval $[0,x_k]$.

\subsection{Relation to the Park method}

Factors (8) are appropriate for describing the behavior of functions in the
vicinity of critical points. A value $x=x_c$ is termed a critical point of a
function $f(x)$ if at this point the latter is either zero, $f(x_c)=0$, or
possesses an algebraic singularity, that is $f(x)\sim (1-x/x_c)^{-\bt}$ as 
$x\ra x_c$, where $\bt$ is positive. This fact was, actually, employed by Park
[33] who suggested a method for locating the critical points and calculating
the critical exponents. His method is formulated as follows. Assume that:
(i) a real function $f(x)$ has a critical point $x_c$; (ii) in the
neighbourhood of the critical point, the function can be represented as 
$$
f(x) \simeq f_0\prod_p (1-B_p\; x)^{-\bt_p} \qquad (x\ra x_c) \; , 
$$
with all $B_p$ and $\bt_p$ being real; (iii) the physical critical point
corresponds to that which is the closest to the origin, such that, arranging 
$B_p$ in the descending order of their absolute values, $|B_p|>|B_{p+1}|$,
one has $x_c=B_1^{-1}$. Then, defining the coefficients $b_n$ by the
expansion 
$$
\ln f(x) = \sum_{n=1}^\infty b_n\; x^n \; , 
$$
one obtains 
$$
B_1 = \lim_{n\ra\infty}\; \frac{b_n}{b_{n-1}} \; , \qquad \bt_1 = 
\lim_{n\ra\infty} \; n\; \frac{b_{n-1}^n}{b_n^{n-1}} \; . 
$$
The proof of this statement is based on the generalized P\'olya theorem
[34], which extends the theorem by P\'olya [35], initially proved for entire
functions of genus zero, when all $\bt_p=-1$, to the case of real $\bt_p$.
The Park method for defining the critical points and critical indices is
closely related to the Pad\'e analysis of logarithmic derivatives of a
series, though in the latter case there is no prior knowledge for the
convergence of estimates from Pad\'e approximants [36], which would be
analogous to the generalized P\'olya theorem.

The principal difference of our approach from the Park method is in the
following. First of all, we never require that the sought function be
exactly factorizable, but we derive the form (9) as an approximation to this
function. Second, we do not impose a constraint that the function must
necessarily possess a critical point, and if so, we consider the function
not solely in the neighbourhood of the latter, but in the whole region $%
[0,x_c]$. Third, since we deal with a much more general case, the amplitudes 
$A_{kp}$ and exponentials $n_{kp}$ are not compulsory real, but may be
complex valued.

\section{Examples}

Any function from the class ${\cal R}$ can be reproduced, according to 
Theorem, exactly, provided that there are enough terms in series (2). 
However, a reproducible function may be not exactly reproduced when the 
asymptotic expansion (2) contains not enough terms. In other words, a 
function to be recovered may be exactly reproducible in principle, but 
in  practice, we may have access to only a few terms in the asymptotic 
expansion. Then an important question to ask is how well the factor 
approximants are able to approximate such a function and also it is 
interesting to observe how the factor approximants converge to the exact 
result. This problem will be considered in the next subsection.

Those functions that are not from the class ${\cal R}$ cannot be reproduced
exactly, but they can be very well approximated by the self-similar factor
approximants (9), as we illustrate in the following subsections.

\subsection{Convergence to exact result}

Consider the function
\be
\label{13}
f(x) = (1+2x)^{3/2}(1+x)^{1/2}(1+0.5x)^{1/3}(1+0.1x)^{1/4}\; ,
\ee
which is from the class ${\cal R}$. Its expansion of eighth order has the 
coefficients $f_0=1$ and
$$
a_1=3.692\; , \qquad a_2=3.521\; , \qquad a_3=0.410 \; , \qquad
a_4=0.025\; , $$
$$
a_5=-0.091\; , \qquad a_6=0.145 \; , \qquad a_7=-0.220\; , \qquad 
a_8=0.335 \; ,
$$
whose behaviour is rather irregular. If we take into account only four 
terms of series (2), then we get the approximant $f_2^*(x)$, with
$$
A_{21}=1.986\; \qquad A_{22}=0.721\; , \qquad n_{21}=1.562\; , \qquad
n_{22}=0.818 \; .
$$
And for the factor approximant $f_3^*(x)$, constructed by means of series 
(2) of sixth order, we find
$$
A_{31}=2\; , \qquad A_{32}=0.960\; , \qquad A_{33}=0.321 \; ,
$$
$$
n_{31} = 1.503\; , \qquad n_{32}=0.583\; , n_{33}=0.398 \; .
$$
The best Pad\'e approximant that can be built of the sixth-order series 
(2) is $P_{[4/2]}(x)$, whose accuracy is compared with that of $f_3^*(x)$ in
Fig. 1. As is evident, the factor approximant is essentially more accurate, 
although is not yet exact. But in the next order, we obtain $f_4^*(x)$, with
$$
A_{41}=2\; , \qquad A_{42}=1\; , \qquad A_{43}=0.5 \; , \qquad
A_{44}=0.1 \; ,
$$
$$
n_{41} =\; \frac{3}{2}, \qquad n_{42}=\frac{1}{2}\; , 
n_{43}= \frac{1}{3} \; , \qquad n_{44} =\frac{1}{4} \; ,
$$
which coincides with the exact function (13).

\subsection{Combination of functions from ${\cal R}$ with exponentials}

The combination of functions from ${\cal R}$ and exponentials are
approximated with a very good accuracy. As an example, let us consider 
\be
\label{14} 
f(x) =\left ( \frac{1+Ax+Bx^2}{1+Cx+Dx^2} \right )^m \; \exp(-x) \; , 
\ee
The choice of the coefficients and the power $m$ is not important, since 
the factor multiplying the exponential pertains to the class ${\cal R}$ 
of exactly reproducible functions. For concreteness, let us take $A=0.5,\;
B=0,\; C=0.5,\; D=0.1$, and $m=0.5$. Then, in the asymptotic series (2), we
have $f_0=1$ and 
$$
a_1=-1.750\; , \qquad a_2 = 2.419\; , \qquad a_3=-3.659\; , 
$$
$$
a_4=6.060\; , \qquad a_5 =-10.499\; , \qquad a_6=18.622\; . 
$$
The first-order approximant (9) is, clearly, too simple for providing a 
good approximation for complicated functions. Therefore, here and in what
follows, we start the analysis with the second-order approximant. For the
considered case, we get in the second order 
$$
A_{21}=1.976\; , \qquad A_{22} =-0.077\; , \qquad n_{21}=-0.471\; , \qquad
n_{22}=10.651 
$$
and in the third order 
$$
A_{31}=1.945\; , \qquad A_{32} =0.501\; , \qquad A_{33}=0.000986\; , 
$$
$$
n_{31}=-0.500\; , \qquad n_{32} =0.497\; , \qquad n_{33}=-1039\; . 
$$
Both approximants $f_2^*(x)$ and $f_3^*(x)$ perfectly reproduce function
(14).

For comparison, we construct the Pad\'e approximants based on the same
number of terms in the series (2). Among all possible Pad\'e approximants 
$P_{[M/N]}(x)$, we select the most accurate for the case considered. Note
that the best Pad\'e approximants are not necessarily diagonal. Here, these
are $P_{[1/5]}(x)$ and $P_{[2/4]}(x)$. The percentage errors of these
approximants, together with the error of $f_3^*(x)$ are shown in Fig. 2. One
can observe that the accuracy of $f_3^*(x)$ is incomparably higher than that
of the best Pad\'e approximants, whose errors grow fast with $x$, reaching
amplitudes of the order of $100\%$. In addition, the approximant $%
P_{[2/4]}(x)$ becomes negative for $x>5$, which is qualitatively wrong for
the positive function (14).

\subsection{Exponential multiplied by functions not from the class ${\cal 
R}$}

As an example of a function having no factors from the class ${\cal R}$, let
us consider 
\be
\label{15} 
f(x) ={\rm tanh}(x) \exp(-x) \; . 
\ee
In its asymptotic series (2), we have $f_0=1$ and 
$$
a_1 = -1 \; , \qquad a_2 =\frac{1}{6} \; , \qquad a_3 = \frac{1}{6}\; , 
$$
$$
a_4 = \frac{1}{120} \; , \qquad a_5 =- \; \frac{31}{360}\; , \qquad a_6 = 
\frac{1}{5040}\; . 
$$
For the second-order approximant (9), we find 
$$
A_{21} = -0.350 + 0.587\; i \; , \qquad A_{22}= A_{21}^* \; , \qquad
n_{21}=0.036 + 0.831\; i \; , \qquad n_{22}=n_{21}^* \; . 
$$
And for the third order, we obtain 
$$
A_{31} = -0.00337 + 0.650\; i \; , \qquad A_{32}= A_{31}^* \; , \qquad
A_{33} = 0.071 \; , 
$$
$$
n_{31}=-0.871 + 0.023\; i \; , \qquad n_{32}=n_{31}^* \; , \qquad
n_{33}=-13.777 \; . 
$$
Again, we compare $f_3^*(x)$ with the most accurate Pad\'e approximants that
for this case are $P_{[1/5]}(x)$ and $P_{[2/4]}(x)$. The corresponding
percentage errors are shown in Fig. 3. Again, we see that $f_3^*(x)$ has
much higher accuracy than the best Pad\'e approximants. The most accurate of
the latter, $P_{[1/5]}(x)$, becomes negative for $x>6$, which is
qualitatively wrong.

A combination of a logarithm and an exponential yields a non-monotone
function 
\be
\label{16} 
f(x) =\ln(1+x)\; \exp(-x) \; . 
\ee
The coefficients of the related asymptotic expansion (2) are $f_0=1$ and 
$$
a_1 = -\; \frac{3}{2} \; , \qquad a_2 =\frac{4}{3} \; , \qquad a_3 = -1\; , 
$$
$$
a_4 = \frac{89}{120} \; , \qquad a_5 =- \; \frac{83}{144}\; , \qquad a_6 = 
\frac{593}{1260}\; . 
$$
For the parameters of the factor approximant $f_2^*(x)$, we get 
$$
A_{21} = 0.930 \; , \qquad A_{22}= 0.013 \; , \qquad n_{21}= -0.466 \; ,
\qquad n_{22}= -83.334 \; , 
$$
and for those of $f_3^*(x)$, we find 
$$
A_{31} = 0.973 \; , \qquad A_{32}= 0.574 \; , \qquad A_{33} = 0.00212 \; , 
$$
$$
n_{31}= -0.363 \; , \qquad n_{32}= -0.215 \; , \qquad n_{33}=-483.556 \; . 
$$
The most accurate Pad\'e approximant here is $P_{[1/5]}(x)$. The
corresponding percentage errors are presented in Fig. 4, from where it is
seen that $f_3^*(x)$ is significantly more accurate than $P_{[1/5]}(x)$.

\subsection{Functions not from the class ${\cal R}$ which converge to a
constant at infinity}

The previous functions converge to zero at infinity. Let us now consider 
the function 
\be
\label{17} 
f(x) =\exp\left ( 1 - \; \frac{1}{\sqrt{1+x}}\right ) \; , 
\ee
which increases at infinity to a finite value. In the expansion (2), we have 
$f_0=1$ and 
$$
a_1=0.5\; , \qquad a_2=-0.25\; , \qquad a_3=0.146 \; , \qquad a_4=-0.091\; , 
$$
$$
a_5=0.059\; , \qquad a_6=-0.038\; , \qquad a_7=0.025 \; , \qquad
a_8=-0.016\; . 
$$
For the second-order factor approximant, we find 
$$
A_{21}=0.570\; , \qquad A_{22}=1.097\; , \qquad n_{21}=-0.671 \; , \qquad
n_{22}=0.805\; , 
$$
and for $f_3^*(x)$, we have 
$$
A_{31}=1.041\; , \qquad A_{32}=0.794\; , \qquad A_{33}=0.265 \; , 
$$
$$
n_{31}=1.243 \; , \qquad n_{32}=-0.922\; , \qquad n_{33}=-0.235 \; . 
$$
The best Pad\'e approximant $P_{[2/2]}(x)$ is less accurate than $f_3^*(x)$, 
as is shown in Fig. 5. The accuracy becomes even better for $f_4^*(x)$,
for which 
$$
A_{41}=0.151\; , \qquad A_{42}=1.023\; , \qquad A_{43}=0.881 \; , \qquad
A_{44}=0.516 \; , 
$$
$$
n_{41}=-0.150 \; , \qquad n_{42}=1.675 \; , \qquad n_{43}=-1.195 \; \qquad
n_{44}=-0.267\; . 
$$
Notice that, passing to higher approximations, the sum $\sum_p n_{kp}$
decreases, 
$$
n_{21}+n_{22} =0.133 \; , \qquad n_{31}+n_{32}+n_{33}=0.086\; , 
$$
$$
n_{41} + n_{42} + n_{43} + n_{44} = 0.063 \; , 
$$
which is the correct trend, since in the limit it should be 
$$
\lim_{k\ra\infty} \; \sum_{p=1}^k \; n_{kp} = 0 \; . 
$$
The accuracy of the factor approximants can also be improved by assuming the
validity of the condition $\sum_p n_{kp}=0$ for finite $k$.

\subsection{Classical example of $\phi^4$ theory with strongly
divergent series}

A classical example of a strongly divergent series is provided by the
asymptotic expansion, in powers of a coupling parameter, of generating
functionals in field theory or of partition functions in statistical
mechanics. The generic structure of such divergent expansions is exemplified
by expanding, in powers of the coupling $g$, the generating functional 
\be
\label{18} 
I(g) = \frac{1}{\sqrt{\pi}} \int_{-\infty}^{+\infty} 
\exp\left ( - \vp^2 - g\vp^4 \right ) \; d\vp \; , 
\ee
typical of $\vp^4$ field theory. In the asymptotic series (2), with $g$
instead of $x$, the coefficients are $f_0=1$ and 
$$
a_n = \frac{(-1)^n}{\sqrt{\pi}\; n!} \; \Gamma\left ( 2n + \frac{1}{2}\right
) \; . 
$$
The latter grows with $n\ra\infty$ as $a_n\sim n^n$. Such a series is
divergent for any finite $g$. For the second-order approximant $I_2^*(g)$,
we find 
$$
A_{21} = 19.141 \; , \qquad A_{22}= 4.859 \; , \qquad n_{21}= -0.00862 \; ,
\qquad n_{22}= - 0.120 \; . 
$$
And for $I_3^*(g)$, we obtain 
$$
A_{31} = 31.220 \; , \qquad A_{32}= 13.317 \; , \qquad A_{33} = 3.464 \; , 
$$
$$
n_{31}= -0.000526 \; , \qquad n_{32}= -0.022 \; , \qquad n_{33}= -0.125 \; . 
$$
The most accurate Pad\'e approximant, constructed with the same number of
the asymptotic terms, is $P_{[3/3]}(g)$. This is compared with $I_3^*(g)$ in
Fig. 6.

Again, we see that the accuracy of the factor approximant highly surpasses
that of the Pad\'e approximant. This is especially noticeable for large
couplings $g$, where the Pad\'e approximant completely fails. Really, 
$P_{[3/3]}(g)$ is finite for $g\ra\infty$, which is in contradiction with 
the behavior of $I(g)$ at large $g$, where the integral (17) tends to zero, 
\be
\label{19} 
I(g) \simeq 1.023\; g^{-0.25} \qquad (g\ra \infty) \; . 
\ee
The factor approximants also decrease in the strong-coupling limit as 
\be
\label{20}
I_2^*(g) \simeq 0.806\; g^{-0.129} \; , \qquad
I_3^*(g) \simeq 0.807 g^{-0.148} \qquad (g\ra \infty) \; . 
\ee
Let us emphasize that all parameters of the self-similar factor approximants
are defined only through the weak-coupling expansion. And it looks almost
mysterious that they can reasonably extrapolate the behavior at the
strong-coupling limit. The accuracy of the approximants (9) increases with
their order. Thus, for the approximant $I_4^*(g)$, we arrive at the
parameters 
$$
A_{41}=43.965\; , \qquad A_{42} = 23.064\; , \qquad A_{43} = 10.294 \; ,
\qquad A_{44}=2.677 \; , 
$$
$$
n_{41} = -0.000025\; , \qquad n_{42}=-0.00259\; , \qquad n_{43}=-0.035\; ,
\qquad n_{44} = -0.124 \; . 
$$
In the strong-coupling limit, this yields 
\be
\label{21} 
I_4^*(g) \simeq 0.810\; g^{-0.161} \qquad (g \ra \infty) \; . 
\ee
Note that the diagonal Pad\'e approximants are always finite at infinity,
thus allowing the approximation of only a very narrow class of functions
[37]. Contrary to this, the factor approximants immediately catch the
correct behavior at infinity; their accuracy increasing with the
approximation order. Hence, the self-similar factor approximants, being
based on an asymptotic expansion at zero, can correctly reproduce the
behavior of the sought function for the whole range of its variable,
including the behavior at infinity.

\subsection{Expansion factor of a three-dimensional polymer chain}

As another physical example, let us consider the expansion factor $\al(z)$
for a three-dimensional polymer chain with excluded-volume interaction,
where $z$ is a dimensionless coupling parameter [38,39]. An asymptotic
series of the form (2), derived by means of perturbation theory [38], yields
the coefficients $f_0=1$ and 
$$
a_1=\frac{4}{3} \; , \qquad a_2 = -2.075385396 \; , \qquad 
a_3=6.296879676\; , 
$$
$$
a_4=-25.05725072 \; , \qquad a_5=116.134785\; , \qquad a_6=-594.71663 \; . 
$$
In the strong-coupling limit, it has been established numerically [37,38]
that $\al(z)$ can be accurately represented by 
\be
\label{22} 
\al(z) \simeq 1.531\; z^{0.3544} + 0.184\; z^{-0.5756} \qquad 
(z \ra \infty) \; . 
\ee
For the approximant $\al_2^*(z)$, we have 
$$
A_{21} = 6.064 \; , \qquad A_{22} = 2.962 \; , \qquad n_{21} = 0.105\; ,
\qquad n_{22} =0.235 \; , 
$$
with the strong-coupling limit 
\be
\label{23} 
\al_2^*(z) \simeq 1.560\; z^{0.340} + 0.151\; z^{-0.660} \qquad 
(z \ra\infty) 
\ee
being very close to the exact numerical value. In the next order, the
parameters of $\al_3^*(z)$ are 
$$
A_{31}=7.019\; , \qquad A_{32}=4.635\; , \qquad A_{33}=2.262\; , 
$$
$$
n_{31}=0.033\; , \qquad n_{32}=0.164 \; , \qquad n_{33}=0.151\; . 
$$
And the strong-coupling limit gives 
\be
\label{24} 
\al_3^*(z) \simeq 1.551\; z^{0.348} + 0.166\; z^{-0.652} 
\qquad (z \ra \infty) \; . 
\ee
Both the coefficients as well as the powers of the strong-coupling
divergence are very accurate, as compared to the numerically found behavior
of $\al(z)$, with the percentage error of $1\%$. The best Pad\'e approximant
here is $P_{[3/3]}(z)$. In Fig. 7, the percentage errors of $\al_3^*(z)$ and 
$P_{[3/3]}(z)$ are shown, compared with the numerically fitted [39] equation 
\be
\label{25} 
\al(z) = \left ( 1 + 7.524\; z + 11.06\; z^2\right )^{0.1172} \; . 
\ee
As is evident, the factor approximant dramatically outperforms the Pad\'e
approximant. The latter completely fails at large $z\ra\infty$, where it is
finite, while $\al(z)$ diverges. To the contrary, the factor approximant 
$\al_3^*(z)$ possesses the correct behavior in the strong-coupling limit.
Moreover, not only the main asymptotic, as $z\ra\infty$, term is very
accurate, but the next term, describing the so-called correction to scaling,
is also of good accuracy. Of special interest is the strong-coupling
exponent 
\be
\label{26} 
\nu \equiv \frac{1}{2} + \frac{1}{4}\; \lim_{z\ra\infty} \; 
\frac{\ln\al(z)}{\ln z} \; , 
\ee
which is a kind of a critical index for the polymer chain [40]. According to
the Muthukumar and Nickel [38,39] numerical estimate, $\nu_{MN}=0.5886$. For
the self-similar factor approximants, we have 
$$
\nu_2^* = 0.585\; , \qquad \nu_3^*=0.587 \; . 
$$
Recent numerical estimates [40,41] give the value $\nu=0.5877\pm 0.0006$,
which is very close to $\nu_3^*$.

\subsection{Nonlinear differential equation for state-dependent
solid friction}

Let us now demonstrate how the self-similar factor approximants can be
employed for solving nonlinear differential equations. Let us consider the
Ruina-Dieterich law of solid friction between two solid surfaces sliding
against each other (see [42,43]). The Ruina-Dieterich law involves the
so-called state variable denoted here as $f$ in a dimensionless form, which 
is usually thought to quantify the true area of contacts of the asperities 
of two solid surfaces. The state variable $f$ obeys the following simple
non-linear differential equation (put in dimensionless form both for $f$ 
and $t$) 
\be
\label{27} 
\frac{df}{dt} = \bt - f^{1-m} \; , 
\ee
where $\bt$ and $m$ are parameters. Equation (27) possesses two
qualitatively different types of solutions corresponding to $m<1$ and $m>1$.

Consider, first, the case $m<1$. Let $m=0.85$, $\bt=0.526$, and the initial
condition $f(0)=0.5$. Equation (27) allows us to derive the short-time
expansion for $f(t)$, presented as the series (2) in powers of time $t$,
simply by using the Taylor expansion formula and by taking successive
derivatives of (27). This gives $f_0=f(0)$ and 
$$
a_1=-0.750\; , \qquad a_2=0.102 \; , \qquad a_3=0.012\; , 
$$
$$
a_4=0.00226\; , \qquad a_5=0.000455\; , \qquad a_6=0.0000861\; . 
$$
By the standard procedure, we obtain the factor approximant $f_2^*(t)$, with 
$$
A_{21}=-0.391 + 0.116\; i \; , \qquad A_{22} = A_{21}^* \; , \qquad n_{21}=
0.682 + 0.932\; i \; , \qquad n_{22}=n_{21}^* \; . 
$$
And the parameters for $f_3^*(t)$ are 
$$
A_{31}= -0.430 + 0.118 \; i \; , \qquad A_{32}=A_{31}^* \; , \qquad A_{33} =
-0.149 \; , 
$$
$$
n_{31} = 0.790 + 0.471\; i \; , \qquad n_{32} = n_{31}^* \; , \qquad n_{33}
= -0.272 \; . 
$$
The accuracy of the approximants can be checked against the direct numerical
solution of Eq. (27). In Fig. 8, we compare the percentage errors of 
$f_3^*(t)$ with those of the best Pad\'e approximants $P_{[2/2]}(t)$ and 
$P_{[3/3]}(t)$. As we see, the factor approximant is considerably more
accurate.

\section{Critical Behavior}

Solutions to physical problems often display a critical behavior, when a
function $f(x)$ tends, at a critical point $x_c$, either to zero or to
infinity. Self-similar factor approximants make it possible to describe
these two types of critical behavior, providing accurate estimates for the
critical points as well as for critical indices. If one is interested solely
in the critical behavior, then our approach reduces to the Park method, as
is discussed in Sec. II. However, we would like to stress that the factor
approximants not solely describe well the critical neighbourhood, but
provide an accurate approximation in the whole region $[0,x_c]$. It is also
important to realize that, in order to achieve an accurate description, one
does not necessarily require the knowledge of a large number of terms in an
asymptotic series, but only a few terms are often sufficient. This applies
to both types of critical points, either zeros or singularities.

\subsection{Critical behavior near zero}

An example of the first type is the function 
\be
\label{28} 
f(x) =\left ( \cos\sqrt{x}\right )^{1/3} \; , 
\ee
tending to zero at the critical point $x_c=\pi^2/4=2.467$ with the critical
index $1/3$. the coefficients of the asymptotic series (2) are $f_0=1$ and 
$$
a_1=-0.167\; , \qquad a_2=-0.014\; , \qquad a_3=-0.00355\; , 
$$
$$
a_4=-0.000982\; , \qquad a_5=-0.000295\; , \qquad a_6=-0.0000937 \; . 
$$
Notice that all coefficients here are negative. Such asymptotic series with
all coefficients of the same sign are known to be very difficult for any
kind of resummation procedure [2]. But, with the factor approximants (9),
there is no problem in approximating the sought function in the whole region
from $x=0$ to the critical point. For the approximant $f_2^*(x)$, we get 
$$
A_{21}=-0.405\; , \qquad A_{22}= -0.024 \; , \qquad n_{21}=0.334\; , \qquad
n_{22}=1.333 \; . 
$$
In this approximation, the critical point $x_c=|A_{21}|^{-1}=2.469$ is
already close to the exact value $\pi^2/4=2.467$. The same quality of
results is obtained for the critical index $n_{21}$, which almost coincides
with the exact index $1/3$. For the next approximation $f_3^*(x)$, we find 
$$
A_{31} = -0.405 \; , \qquad A_{32} = -0.044\; , \qquad A_{33}=-0.00541\; , 
$$
$$
n_{31}=0.333\; , \qquad n_{32}=0.375\; , \qquad n_{33}=2.791\; . 
$$
The approximant $f_3^*(x)$ is practically indistinguishable from function
(28) in the whole region $[0,x_c]$. The critical point 
$x_c=|A_{31}|^{-1}=2.467$ yields the exact value $\pi^2/4$ with a
precision of about $10^{-3}$. The corresponding critical index $n_{31}$
differs from the exact index in the seventh decimal digit. The best Pad\'e
approximant $P_{[3/2]}(x)$ is much less accurate, giving the critical point 
$x_c=2.507$ and a very inaccurate critical index equal to $1$ instead of 
$1/3$. The accuracies of $f_3^*(x)$ and of $P_{[3/2]}(x)$ are compared in 
Fig. 9.

\subsection{Critical singular behavior}

In several physical problems [42], the critical behavior is described by the
function 
\be
\label{29} 
f(x) = \frac{\pi}{2{\rm arcos}x} \; , 
\ee
which diverges at the critical point $x_c=1$, with the critical index $1/2$.
The asymptotic series (2) has the coefficients $f_0=1$ and 
$$
a_1=0.637\; , \qquad a_2=0.405\; , \qquad a_3=0.364\; , 
$$
$$
a_4=0.299\; , \qquad a_5=0.281\; , \qquad a_6=0.248\; . 
$$
This is again the case of a series with constant-sign coefficients, which
is a difficult problem for resummation. For the approximant $f_2^*(x)$, 
we find 
$$
A_{21}=-0.990 \; , \qquad A_{22} = 0. 778 \; , \qquad n_{21}=-0.514 \; ,
\qquad n_{22}=0.164 \; . 
$$
This gives the critical point $x_c=|A_{21}|^{-1}=1.010$ and the critical
index $|n_{21}|$ in a very good agreement with the exact values. In the next
order, with an accuracy up to three decimal digits, we obtain 
$$
A_{31}=-1\; , \qquad A_{32} = 0.912 \; , \qquad A_{33}=0.363\; , 
$$
$$
n_{31}=-0.501\; , \qquad n_{32}=0.091 \; , \qquad n_{33}=0.146\; . 
$$
The critical point $x_c=|A_{31}|^{-1}=1$ coincides with the exact value, and
the critical index $|n_{31}|$ is also practically equal to the exact index 
$1/2$. The best Pad\'e approximant, $P_{[4/1]}(x)$, is much less accurate,
yielding the critical point $x_c=1.064$ and a bad estimate for the critical
index, equal to $1$ instead of $1/2$. The accuracy of the approximants 
$f_3^*(x)$ and $P_{[4/1]}(x)$ are compared in Fig. 10.

\subsection{Critical behavior in nonlinear differential equation}

Critical behavior may arise in solutions of differential equations. For
instance, the Ruina-Dieterich law of solid friction, given by Eq. (27), is
qualitatively different for $m<1$ and $m>1$. For $m<1$, the solution has no
zeros. But for $m>1$, the solution becomes zero at a critical time $t_c$
approached with the critical index $1/m$. Let $m=1.5$ and all other
parameters be the same as for the noncritical case considered in the
previous section. Then the critical time is $t_c=0.33004$. For the
short-time expansion in powers of time $t$, we have $f_0=f(0)=0.5$ and 
$$
a_1=-1.776\; , \qquad a_2=-1.256\; , \qquad a_3=-1.706 \; , 
$$
$$
a_4=-3.024\; , \qquad a_5=-6.114\; , \qquad a_6=-13.388 \; . 
$$
The asymptotic series strongly diverges. Constructing the factor approximant 
$f_2^*(t)$, we get 
$$
A_{21}=-3.049\; , \qquad A_{22}=-0.558 \; , \qquad n_{21}=0.616 \; , \qquad
n_{22}=-0.181 \; . 
$$
This gives the critical time $t_c=|A_{21}|^{-1}=0.328$ and the critical
index $n_{21}$, estimating rather well the corresponding exact values 
$t_c=0.330$ and $1/m=0.667$, respectively. For the parameters of $f_3^*(t)$,
we derive 
$$
A_{31}=-3.036\; , \qquad A_{32}=-1.676\; , \qquad A_{33}=-0.225\; , 
$$
$$
n_{31}=0.630\; , \qquad n_{32}=-0.046 \; , \qquad n_{33}=-0.269\; . 
$$
The critical time $t_c=|A_{31}|^{-1}=0.3294$ approximates the exact
numerical value with a good accuracy, the error being only $-0.19\%$. The
related critical index $n_{31}$ is also close to the exact index $0.667$.
The most accurate Pad\'e approximant $P_{[3/3]}(t)$ yields a much worse
approximation, with the critical time $t_c=0.34134$ and the critical index 
$1$ instead of $2/3$. The relative accuracies of the approximants $f_3^*(x)$
and $P_{[3/3]}(x)$ are demonstrated in Fig. 11.

\subsection{Two examples of critical phenomena in statistical physics}

An analysis of critical behavior would not be complete without considering
critical phenomena of statistical mechanics. Consider, for instance, the
three-dimensional spin-$1/2$ Ising model with a simple cubic lattice.
Thermodynamic characteristics of this model can be presented in the form of
high-temperature expansions in powers of the parameter $v \equiv {\rm tanh}
(J/k_BT)$, where $J$ is the exchange integral and $T$, the temperature. For
example, the second derivative of the susceptibility with respect to an
external field 
\be
\label{30} 
\chi_0^{(2)} = \left.\frac{\partial^2\chi}{\partial H^2} \right
|_{H=0} 
\ee
is known [44] as the series expansion containing the powers of $v$ up to 
$v^{17}$. The first coefficients of this expansion are $f_0=-2$ and 
$$
a_1=24\; , \qquad a_2=318\; , \qquad a_3=3240\; , \qquad a_4=28158\; , 
$$
$$
a_5=220680\; , \qquad a_6=1604406\; , \qquad a_7=11029560\; , \qquad
a_8=72559422\; . 
$$
The coefficients are of constant sign and are growing fast. Applying the
method of the factor approximants, we have in the second order 
$$
A_{21}=-4.602\; , \qquad A_{22}=7.373 \; , \qquad n_{21}=-4.300\; , \qquad
n_{22}=0.571 \; , 
$$
where we limit ourselves by three decimal digits. This gives the critical
point $v_c=|A_{21}|^{-1}=0.217$ at which the function diverges with the
critical index $|n_{21}|$. These values can be compared with accurate Monte
Carlo calculations [45] yielding the critical point $v_c=0.218092$ and with
a refined analysis by means of integral approximants [44] giving the
critical index $4.37$. In the third order, again with an accuracy up to
three decimals, we obtain 
$$
A_{31}=-4.601\; , \qquad A_{32}=7.375\; , \qquad A_{33}=-44.934\; , 
$$
$$
n_{31}=-4.301\; , \qquad n_{32}=0.571\; , \qquad n_{33}=0 \; . 
$$
Then the critical point is $v_c=|A_{31}|^{-1}=0.217$ and the critical index
is $|n_{31}|$, which are practically the same as in the second
approximation. The best Pad\'e approximant $P_{[1/4]}(v)$ results in rather
inaccurate values for the critical point $v_c=0.153$ and the critical index
equal to $1$. The fourth-order factor approximant does not lead to a
noticeable change of the critical parameters, as compared to the third-order
one, also giving $v_c=0.217$ and the critical index $4.301$. The percentage
errors of those estimates are about $1\%$.

As another example of high-temperature series expansions, let us consider
such series for the $(2+1)$-dimensional Ising model, defined by the
Hamiltonian 
\be
\label{31} 
H =\sum_i \left ( 1 -\sigma_i^3\right ) - x \sum_{<ij>}
\sigma_i^1\; \sigma_j^1 - h\sum_i \sigma_i^1 \; , 
\ee
in which $\sigma_i^\al$, with $\al=1,2,3$, are Pauli matrices; the index $i$
enumerates the sites on the two-dimensional spatial lattice; $<ij>$ denotes
nearest-neighbour pairs of sites; $x$ is an effective coupling parameter
corresponding to the inverse temperature in the Euclidean formulation of
field theory; and $h$ stands for magnetic field. Here, we consider the
triangular lattice. Let us take, e.g., a series in powers of $x$ for the
mass gap $F$ at zero magnetic field, 
\be
\label{32} 
F \equiv E_1 - E_0 = F(x) \; , 
\ee
which is the difference between the energy of the first excited level and
the ground-state energy. The series (2) for this case [46] has the
coefficients $f_0=2$ and 
$$
a_1=-3\; , \qquad a_2=-3\; , \qquad a_3=-5.25\; , \qquad a_4=-15.75\; , 
$$
$$
a_5=-49.265625\; , \qquad a_6=-173.3554688\; , \qquad a_7=-627.602783\; ,
\qquad a_8=-2397.718506\; . 
$$
For the parameters of the second-order factor approximant $F_2^*(x)$, we
find 
$$
A_{21}=-4.7404\; , \qquad A_{22}=1.7404\; , \qquad n_{21}=0.6582\; , \qquad
n_{22}=0.0691\; . 
$$
The critical point, following from here, is $x_c=|A_{21}|^{-1}=0.2110$. The
mass gap tends to zero at $x_c$ as $F\sim (x_c - x)^\nu$, with the critical
index $\nu=n_{21}$. For the third-order approximant $F_3^*(x)$, we have 
$$
A_{31}=-4.7826\; , \qquad A_{32}=-3.5899\; , \qquad A_{33}=1.5136\; , 
$$
$$
n_{31}=0.6211\; , \qquad n_{32}=0.0458\; , \qquad n_{33}=0.0890\; . 
$$
Hence, the critical point is $x_c=|A_{31}|^{-1}=0.2091$ and the critical
index is $\nu=n_{31}$. In the next order, for the parameters of $F_4^*(x)$,
we obtain 
$$
A_{41}=-4.7629\; , \qquad A_{42}=-2.1880\; , \qquad A_{43}=3.9403\; , \qquad
A_{44}=1.2670\; , 
$$
$$
n_{41}=0.6445\; , \qquad n_{42}=0.0375\; , \qquad n_{43}=0.0007\; , \qquad
n_{44}=0.1178\; . 
$$
This results in the critical point $x_c=|A_{41}|^{-1}=0.2100$ and the
critical index $\nu=n_{41}$. These values can be compared with those
summarized in Ref. [44], where $x_c$ varies between $0.2097$ and $0.2098$,
while the index $\nu$ is between $0.627$ and $0.641$, which is very close to
our results.

\section{Conclusion}

By employing the self-similar approximation theory, we have derived a new
class of approximants, which, because of their form, we call the {\it 
self-similar factor approximants}. All control parameters of this class
of approximants are defined by means of the accuracy-through-order
relationship. These approximants reproduce {\it exactly} a wide class of
functions from the sole knowledge of their asymptotic expansions. For other
functions not from the exactly reproducible class, the factor approximants
provide a very high accuracy, which is essentially higher than that
given by the best Pad\'e approximants constructed with the same number of
asymptotic terms.

The self-similar factor approximants are able to reproduce with a good
accuracy various kinds of functions, diminishing and increasing, monotone 
and non-monotone, on finite or infinite intervals. The asymptotic series,
used for constructing the factor approximants, can have their coefficients
with alternating signs or with constant signs. The series can also be
strongly divergent. Although the factor approximants are derived from 
asymptotic series for a variable in the vicinity of zero, they extrapolate 
well to the behavior of the sought functions at infinity. The approximants 
are able to predict the occurrence of critical phenomena, providing 
accurate values for both the critical points and critical indices.

Note that a natural generalization involves combining the factor
approximants with the self-similar root approximants and exponential
approximants. The latter can be treated as a limiting case of the factor
approximants, since the renormalized factor function (8) tends, as $n_{k_p}%
\ra\infty$, to an exponential. The self-similar exponential approximants, as
has been recently shown [47], enjoy the property of exactly reconstructing
exponential functions. Therefore, the factor approximants, together with
their limiting exponential forms, and being combined with the self-similar
root approximants, provide a powerful tool for an accurate reconstruction of
a very wide class of functions.

At first glance, it may appear mysterious that, knowing solely the behavior
of a function in the vicinity of zero, it is possible to correctly predict
its behavior at infinity or near a critical point. However, there is no
miracle here. The coefficients of an asymptotic series contain a great deal
of information about their parent function, provided that the latter are 
differentiable up to sufficiently high order. Then, the main problem is that 
this information is hidden, encoded. And one needs to possess a guide and 
a key for decoding that information. The idea of {\it group self-similarity} 
for subsequent approximations [17--27] serves as a guide pointing at the 
general properties of the sought function, which allows for the extrapolation 
of the given approximations. And the self-similar factor approximants is 
a practical key for realizing this extrapolation.

Since the self-similar approximation method presented here is capable of
capturing the features of rather complex functions with very good
accuracy from the knowledge of a few numerical coefficients, one can view
this approach as a complexity reduction scheme, or better as an
encoding-decoding scheme. In the language of algorithmic complexity theory 
\cite{algocompl}, the differentiable functions with only isolated critical
points have a low degree of complexity.

\newpage

\begin{center}
{\large {\bf Figure Captions}}
\end{center}

\vskip 1cm

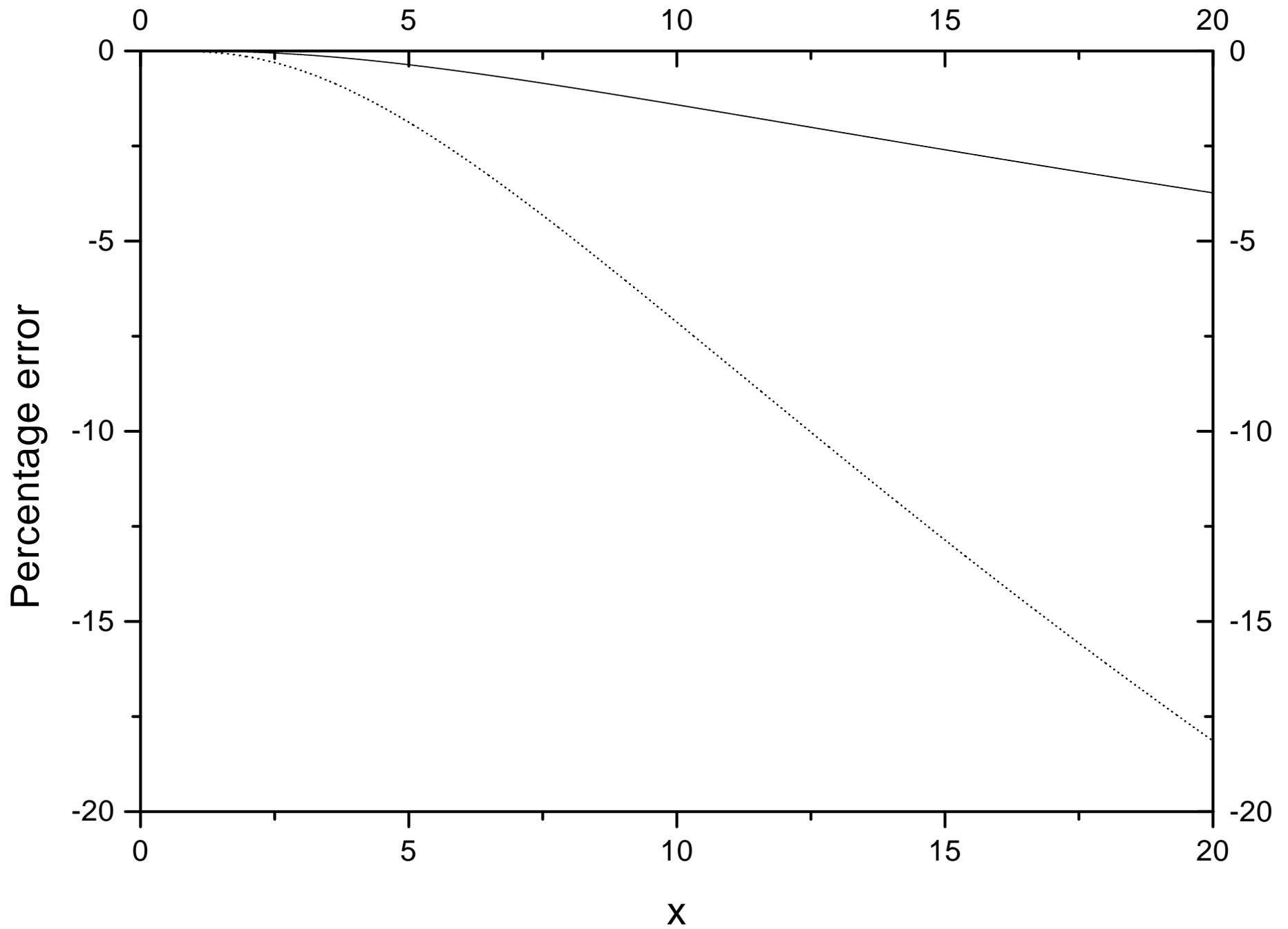
{\bf Fig. 1}. Percentage errors of the self-similar factor approximant
$f_3^*(x)$ (solid line) and of the Pad\'e approximant $P_{[4/2]}(x)$ (dotted
line), as compared with function (13).

\vskip 1cm

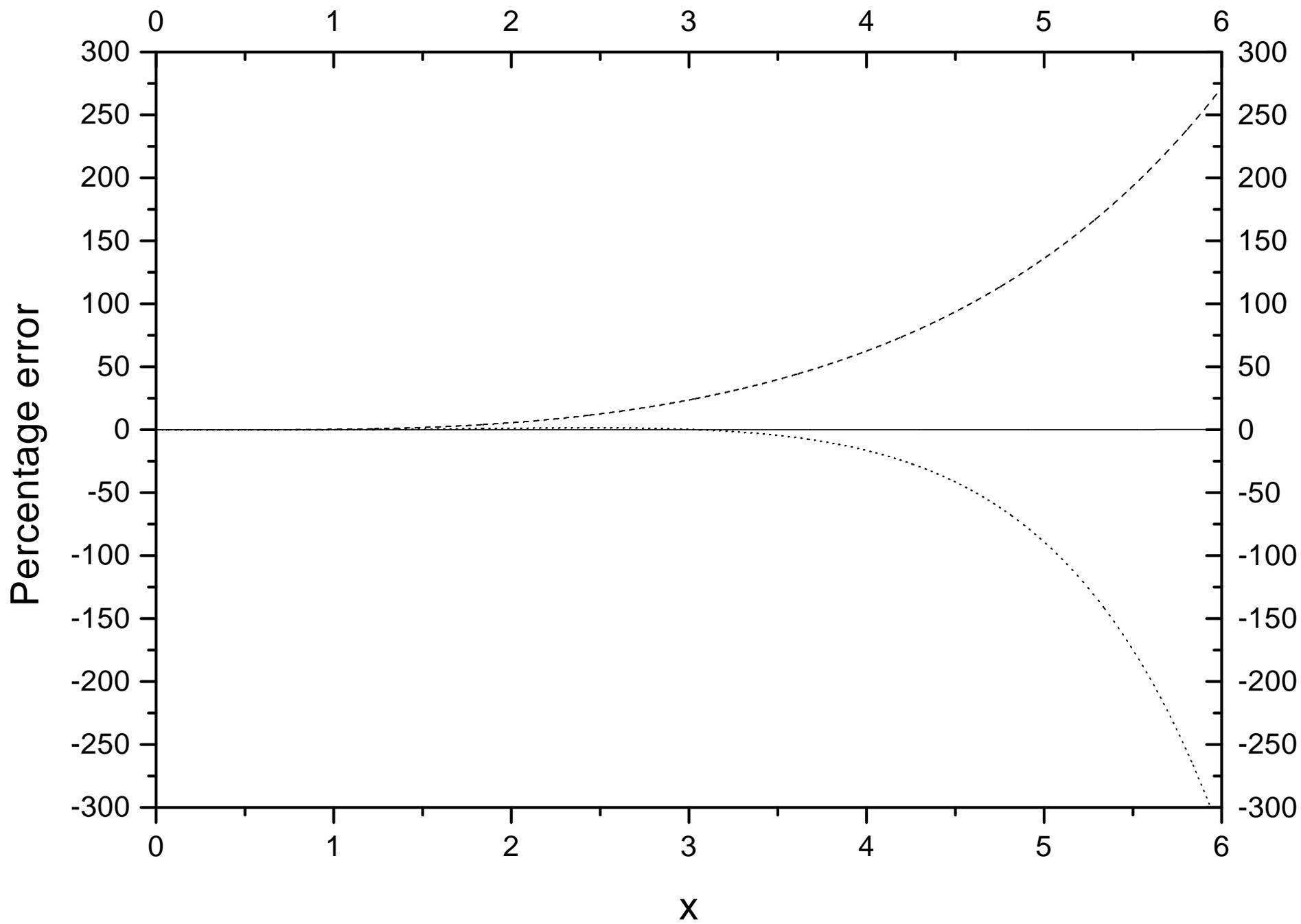
{\bf Fig. 2}. Percentage errors of the self-similar factor approximant 
$f_3^*(x)$ (solid line) and of the Pad\'e approximants $P_{[1/5]}(x)$ (dashed
line) and $P_{[2/4]}(x)$ (dotted line), compared with function (14).

\vskip 1cm

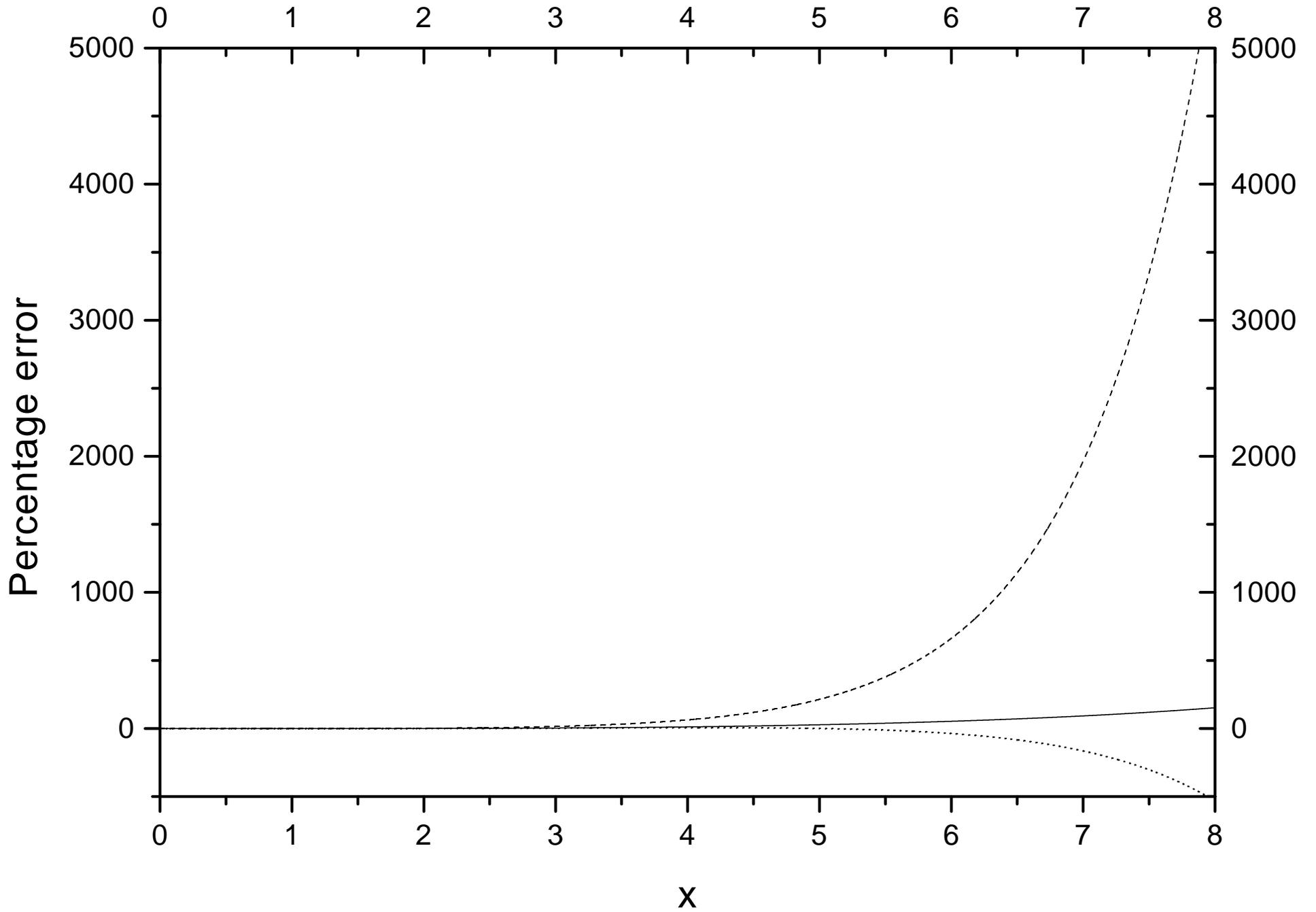
{\bf Fig. 3}. Percentage errors of the factor approximant $f_3^*(x)$ (solid
line) and of the Pad\'e approximants $P_{[1/5]}(x)$ (dotted line) and 
$P_{[2/4]}(x)$ (dashed line), compared with function (15).

\vskip 1cm

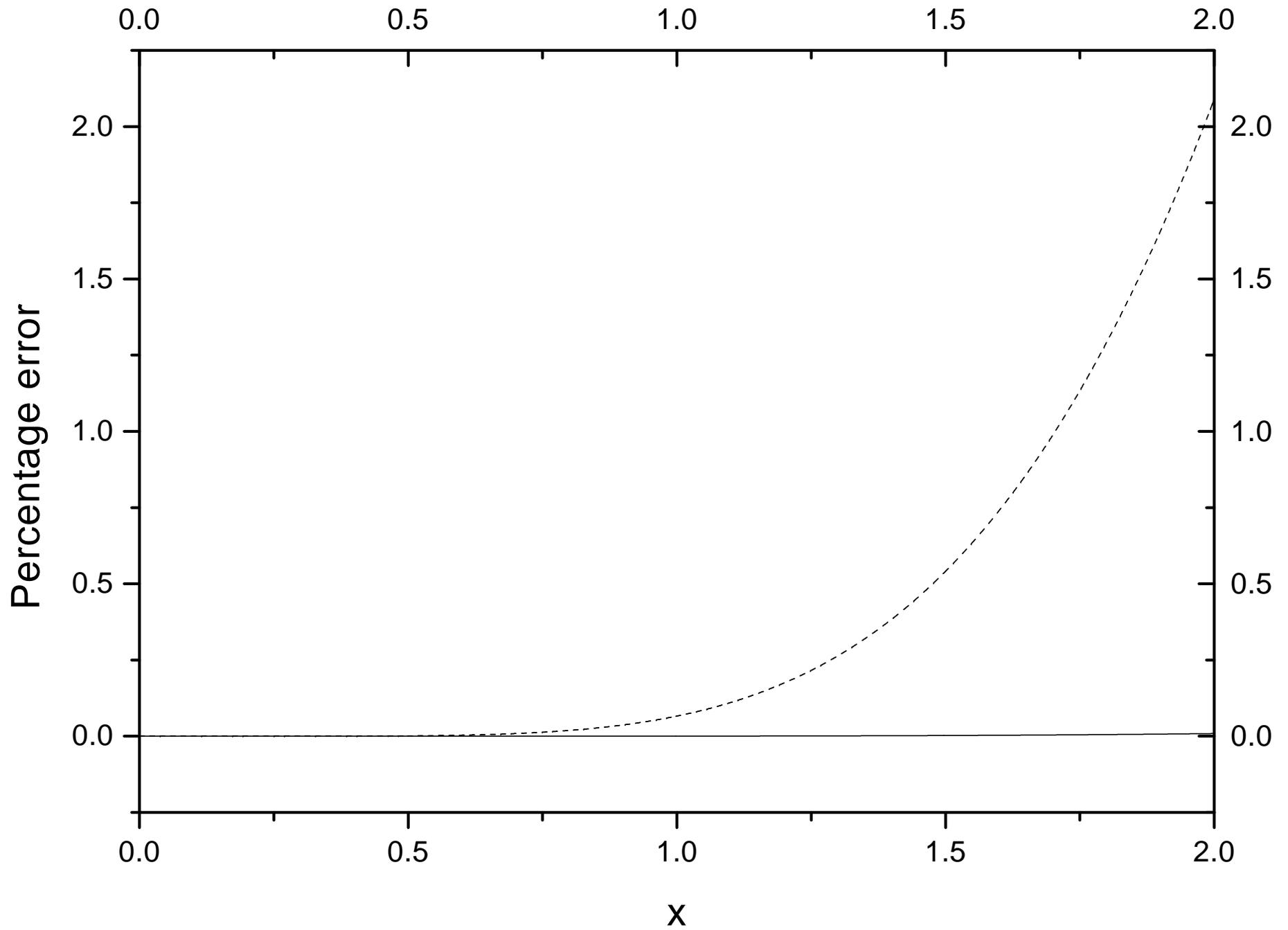
{\bf Fig. 4}. Percentage errors of $f_3^*(x)$ (solid line) and $P_{[1/5]}(x)$
(dashed line), approximating function (16).

\vskip 1cm

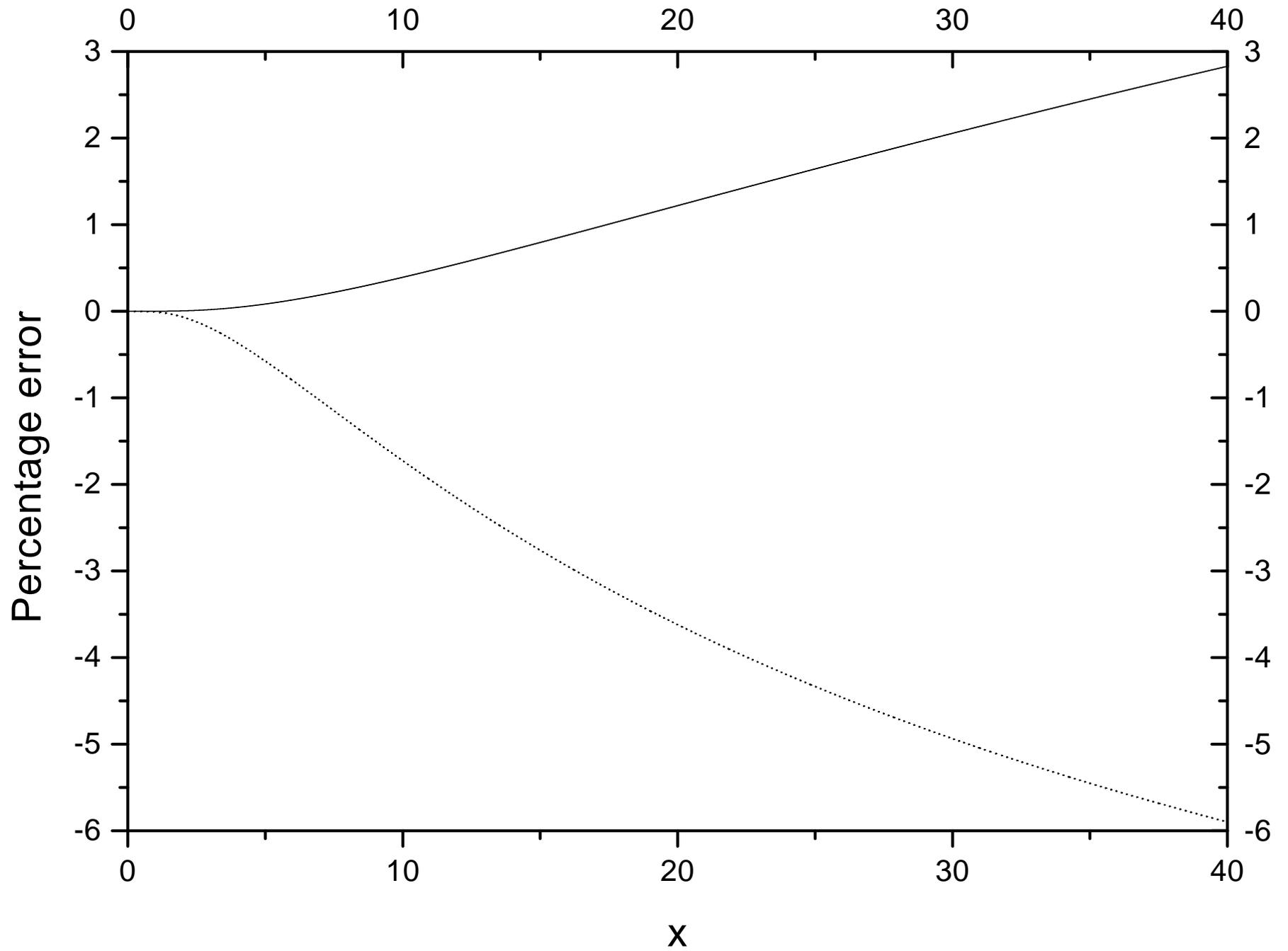
{\bf Fig. 5}. Percentage errors of $f_3^*(x)$ (solid line) and $P_{[2/2]}(x)$
(dotted line), compared with function (17).

\vskip 1cm

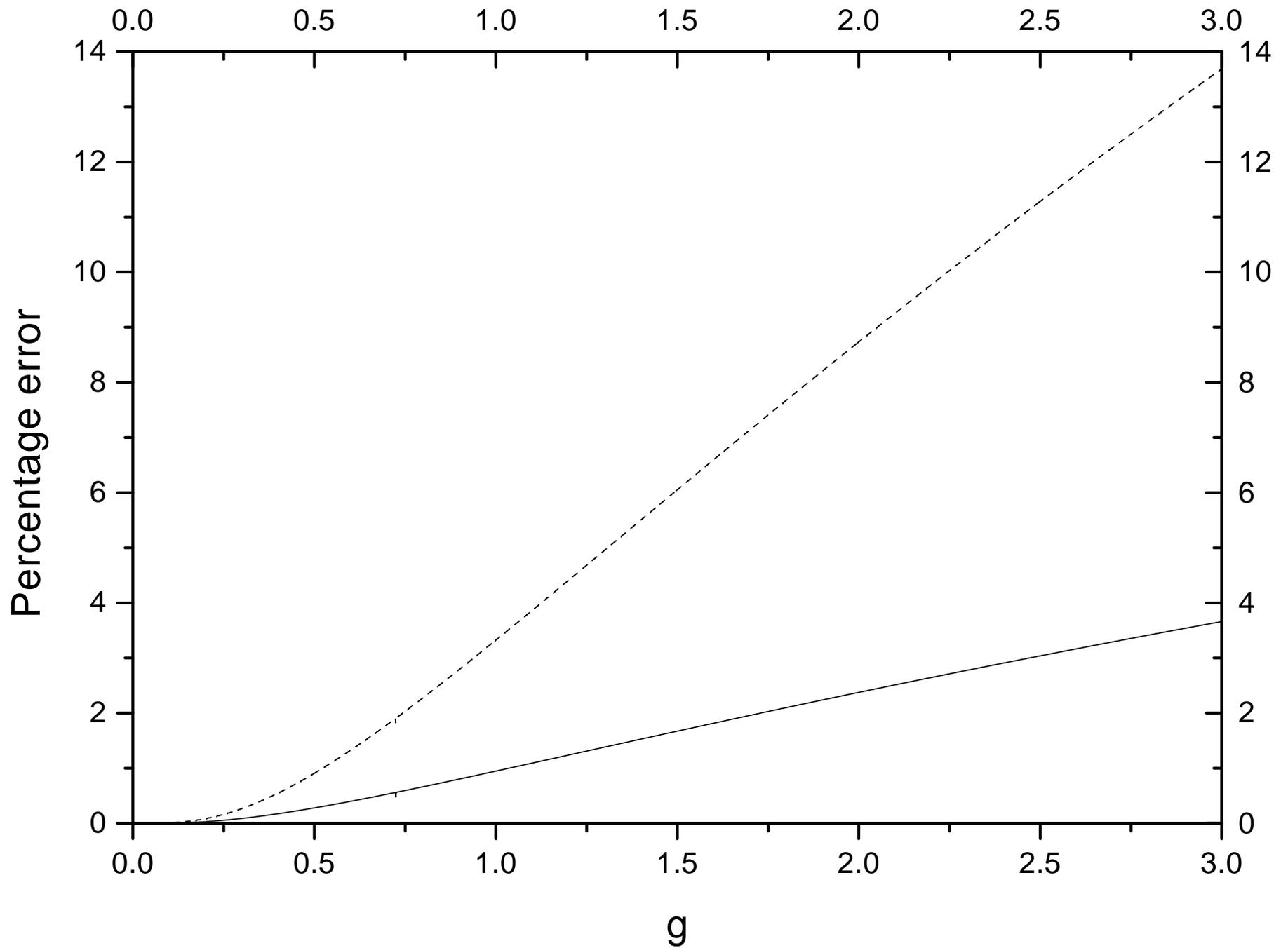
{\bf Fig. 6}. Percentage errors of $I_3^*(g)$ (solid line) and $P_{[3/3]}(g)$
(dashed line), compared with the numerical values of integral (18).

\vskip 1cm

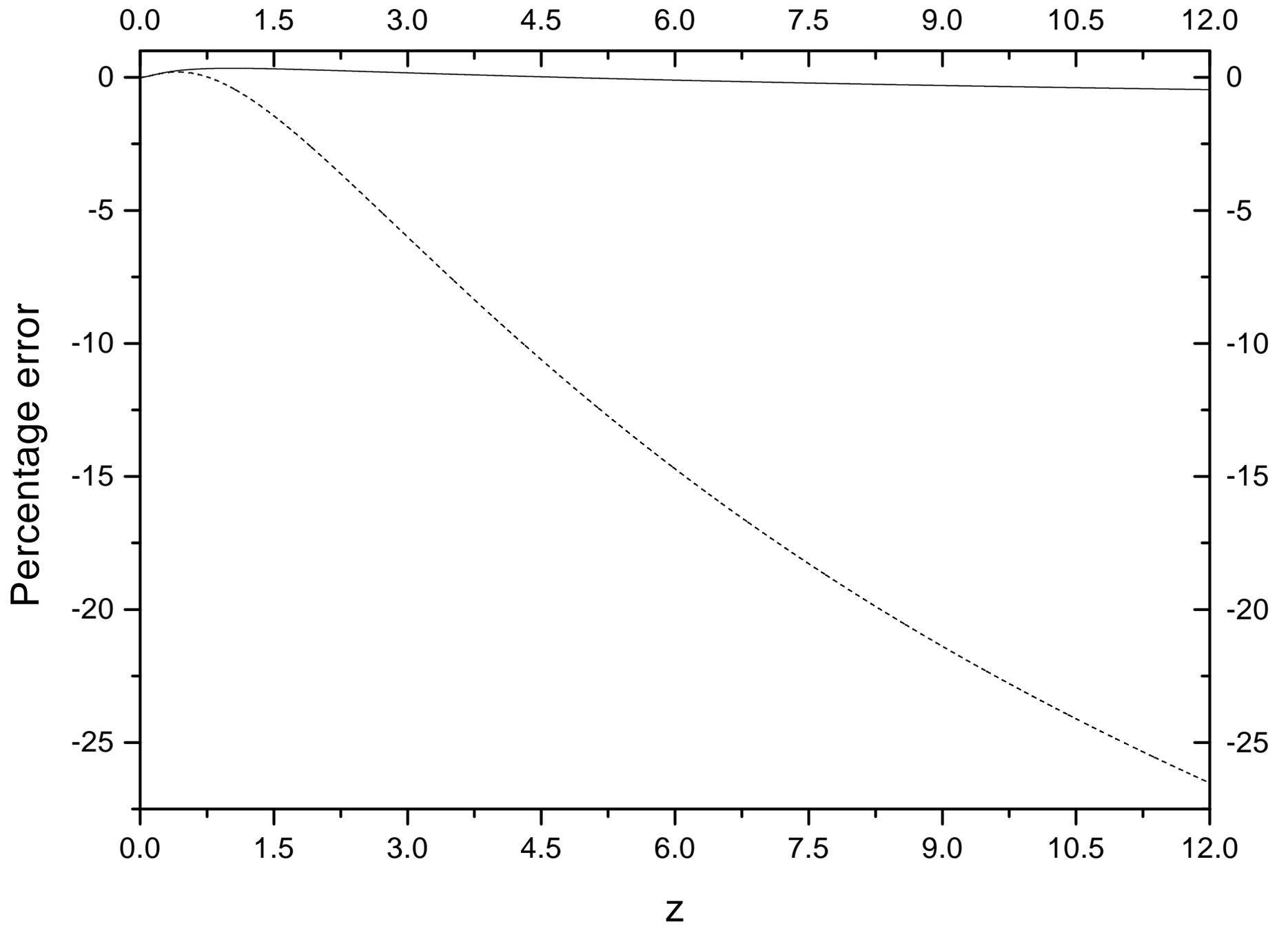
{\bf Fig. 7}. Percentage errors of the approximants for the expansion factor 
$\al_3^*(z)$ (solid line) and $P_{[3/3]}(z)$ (dashed line), compared with
numerical values.

\vskip 1cm

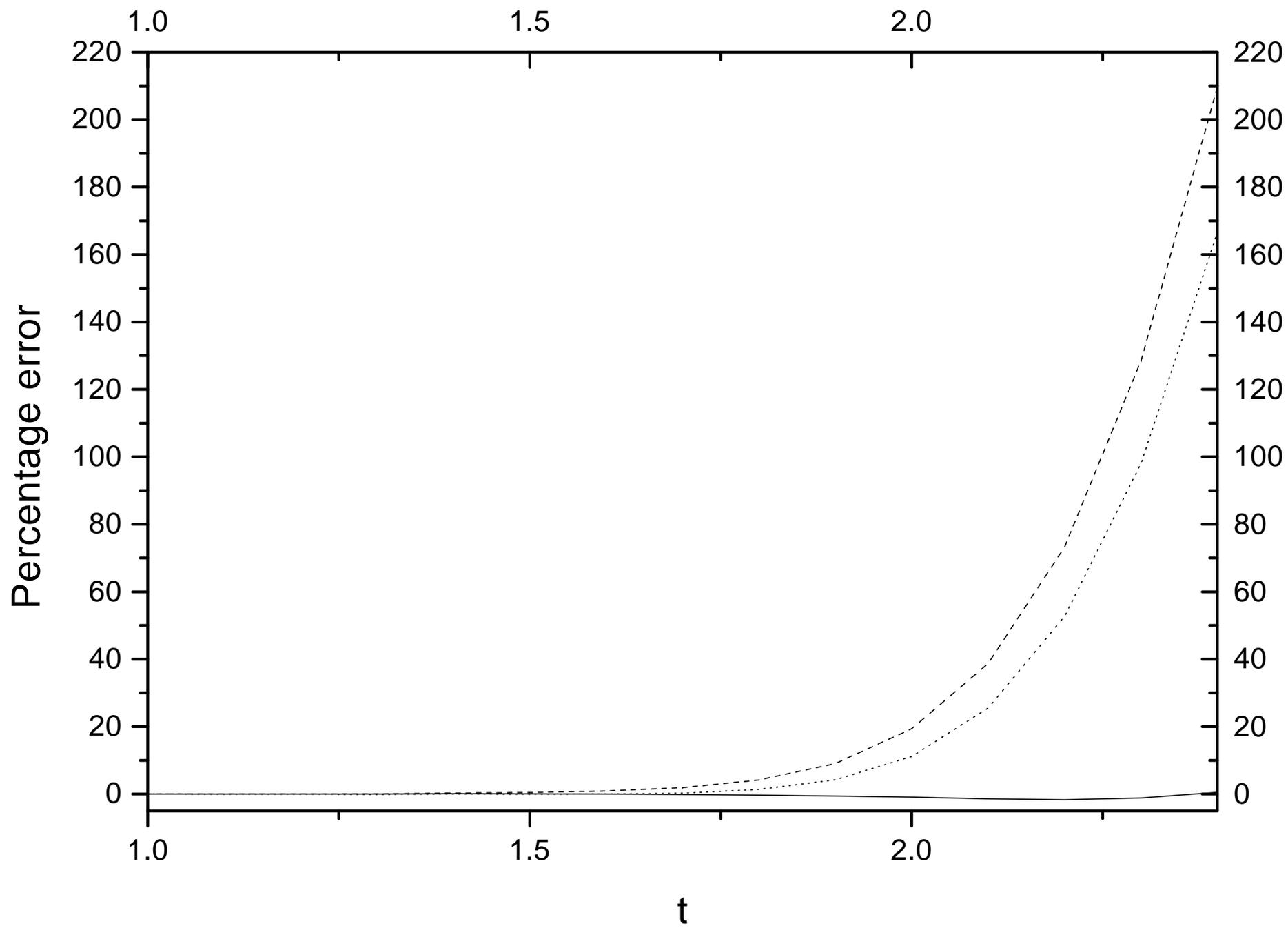
{\bf Fig. 8}. Percentage errors of the approximants for friction, $f_3^*(t)$
(solid line), $P_{[2/2]}(t)$ (dotted line), and $P_{[3/3]}(t)$ (dashed
line), compared with an exact numerical solution of Eq. (27).

\vskip 1cm

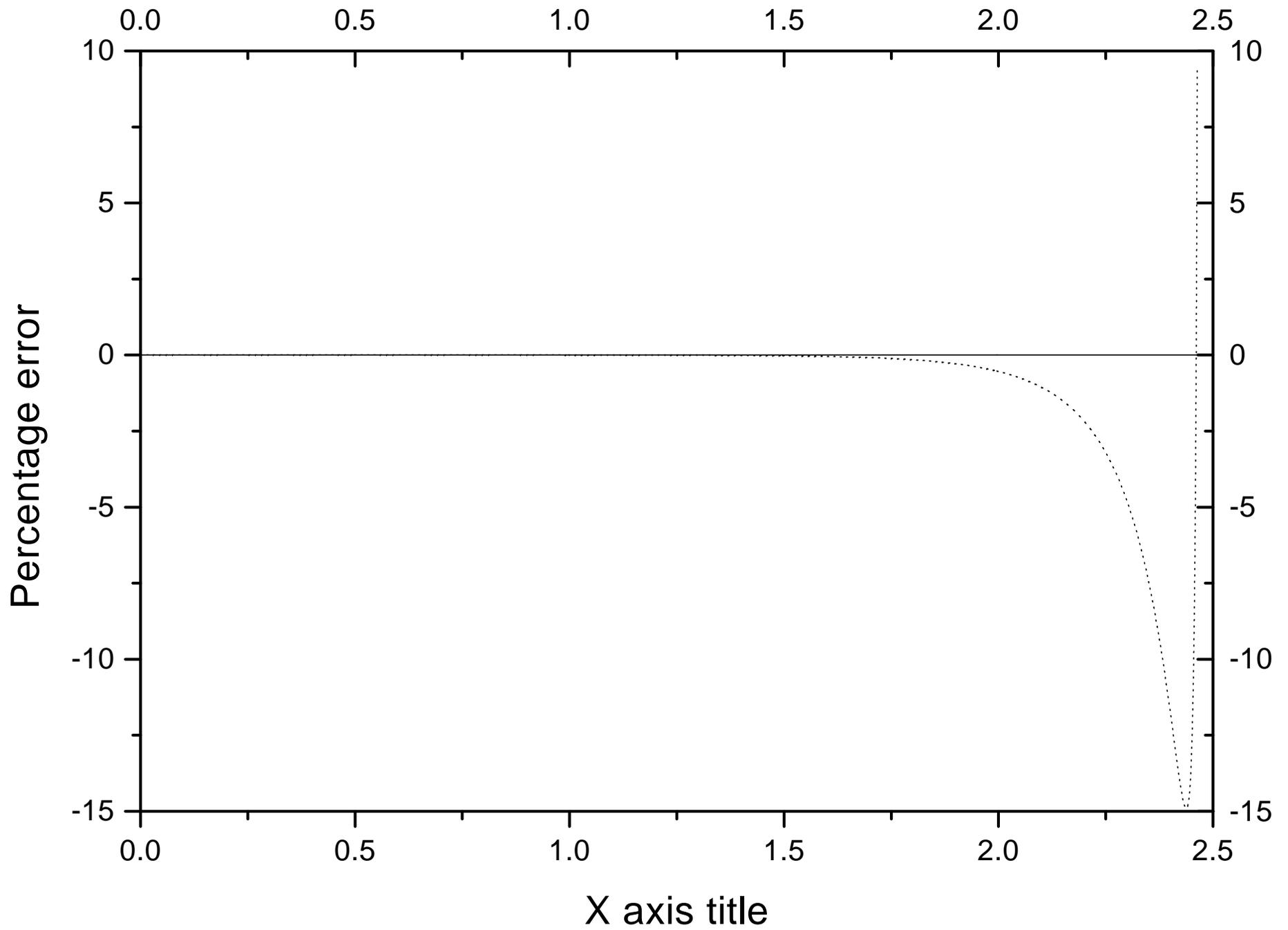
{\bf Fig. 9}. Percentage errors of $f_3^*(x)$ (solid line) and $P_{[3/2]}(x)$
(dotted line) as compared to function (28).

\vskip 1cm

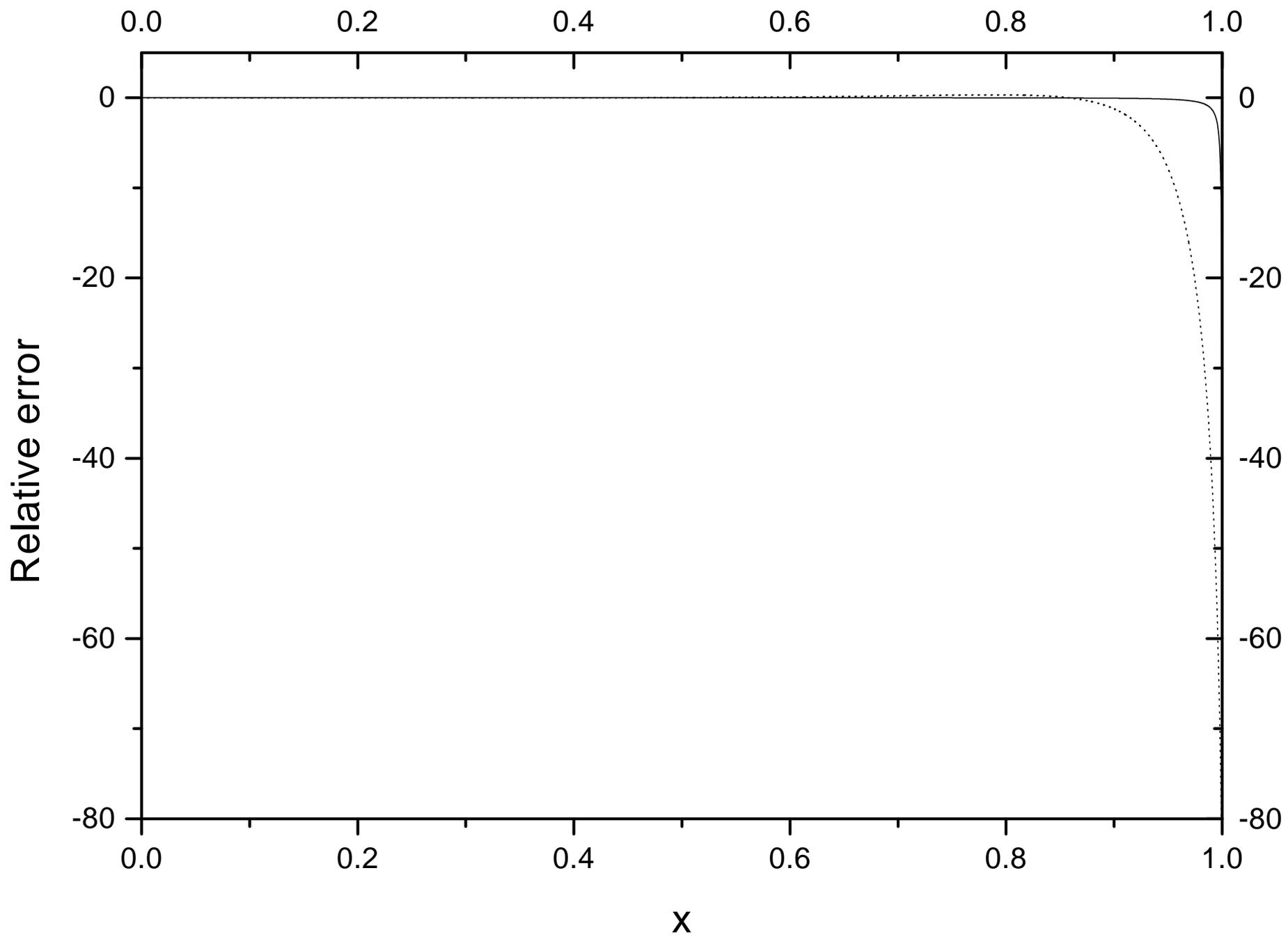
{\bf Fig. 10}. Percentage errors of $f_3^*(x)$ (solid line) and 
$P_{[4/1]}(x)$ (dotted line) with respect to function (29).

\vskip 1cm

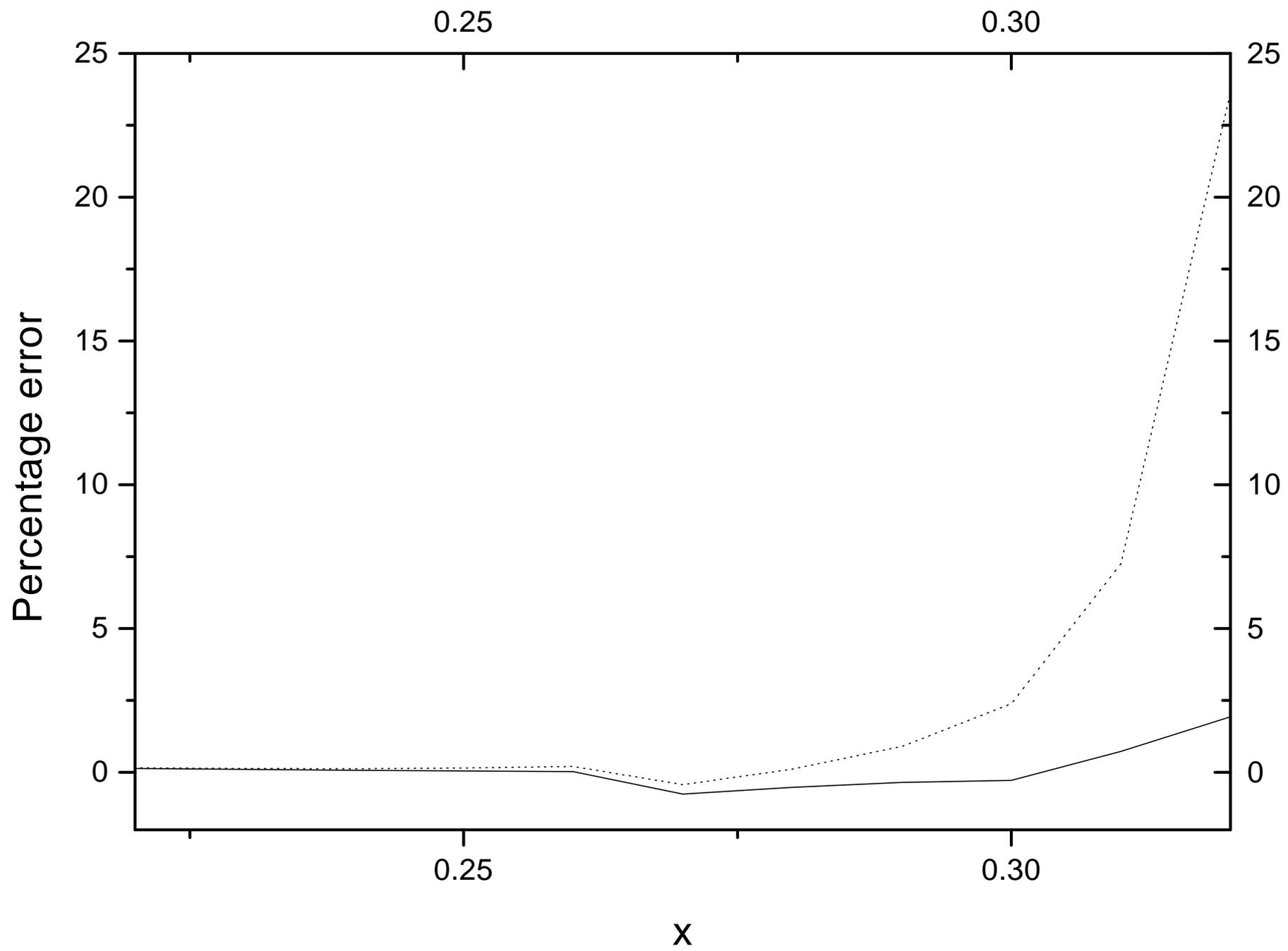
{\bf Fig. 11}. Percentage errors of $f_3^*(x)$ (solid line) and $P_{[3/3]}(x)
$ (dotted line), compared with the numerical solution of Eq. (27).


\newpage

\begin{thebibliography}{99}
\bibitem{1}  G. A. Baker and P. Graves-Moris, {\it Pad\'e Approximants}
(Cambridge University, Cambridge, 1996).

\bibitem{2}  J. Zinn-Justin, {\it Quantum Field Theory and Critical Phenomena%
} (Clarendon, Oxford, 1996).

\bibitem{3}  V. I. Yukalov, Mosc. Univ. Phys. Bull. {\bf 31}, 10 (1976).

\bibitem{4}  V. I. Yukalov, Theor. Math Phys. {\bf 28}, 652 (1976).

\bibitem{5}  V. I. Yukalov, Physica A {\bf 89}, 363 (1977).

\bibitem{6}  W. E. Caswell, Ann. Phys. (N. Y.) {\bf 123} 153 (1979).

\bibitem{7}  I. Halliday and P. Suranyi, Phys. Rev. D {\bf 21}, 1529 (1980).

\bibitem{8}  P. M. Stevenson, Phys. Rev. D {\bf 23}, 2916 (1981).

\bibitem{9}  J. Killingbeck, J. Phys. A {\bf 14}, 1005 (1981).

\bibitem{10}  I. D. Feranchuk and L. I. Komarov, Phys. Lett. A {\bf 88}, 211
(1982).

\bibitem{11}  A. Okopi\'nska, Phys. Rev. D {\bf 35}, 1835 (1987).

\bibitem{12}  J. Honkonen and M. Nalimov, Phys. Lett. B {\bf 459}, 582
(1999).

\bibitem{13}  A. N. Sissakian and I. L. Solovtsov, Phys. Part. Nucl. {\bf 30}%
, 1057 (1999).

\bibitem{14}  T. S. Evans, M. Ivin, and M. M\"obius, Nucl. Phys. B {\bf 577}%
, 325 (2000).

\bibitem{15}  M. B. Pinto and R. O. Ramos, Phys. Rev. D {\bf 61}, 125016
(2000).

\bibitem{16}  P. W. Courteille, V. S. Bagnato, and V. I. Yukalov, Laser
Phys. {\bf 11}, 659 (2001).

\bibitem{17}  V. I. Yukalov, Int. J. Mod. Phys. B {\bf 3}, 1691 (1989).

\bibitem{18}  V. I. Yukalov, Physica A {\bf 167}, 833 (1990).

\bibitem{19}  V. I. Yukalov, J. Math. Phys. {\bf 32}, 1235 (1991).

\bibitem{20}  V. I. Yukalov, J. Math. Phys. {\bf 33}, 3994 (1992).

\bibitem{21}  V. I. Yukalov and E. P. Yukalova, Physica A {\bf 225}, 336
(1996).

\bibitem{22}  V. I. Yukalov and E. P. Yukalova, Ann. Phys. (N. Y.) {\bf 277}%
, 219 (1999).

\bibitem{23}  V. I. Yukalov and S. Gluzman, Phys. Rev. Lett. {\bf 79}, 333
(1997).

\bibitem{24}  S. Gluzman and V. I. Yukalov, Phys. Rev. E {\bf 55}, 3983
(1997).

\bibitem{25}  V. I. Yukalov and S. Gluzman, Phys. Rev. E {\bf 55}, 6552
(1997).

\bibitem{26}  V. I. Yukalov and S. Gluzman, Phys. Rev. E {\bf 58}, 1359
(1998).

\bibitem{27}  S. Gluzman and V. I. Yukalov, Phys. Rev. E {\bf 58}, 4197
(1998).

\bibitem{28}  B. B. Mandelbrot, {\it The Fractal Geometry of Nature}
(Freeman, New York, 1983).

\bibitem{29}  H. Kr\"oger, Phys. Rep. {323}, 81 (2000).

\bibitem{30}  V. I. Yukalov, Mod. Phys. Lett. B {\bf 14}, 791 (2000).

\bibitem{31}  S. Lang, {\it Algebra} (Addison-Wesley, Reading, 1984).

\bibitem{32}  S. Gluzman and D. Sornette, Phys. Rev. E  in press (2002)
cond-mat/0111181.

\bibitem{33}  D. Park, Physica {\bf 22}, 932 (1956).

\bibitem{34}  C. J. Thompson, A. J. Guttman, and B. W. Ninham, J. Phys. C 
{\bf 2}, 1889 (1969).

\bibitem{35}  G. P\'olya, Acta Sci. Math. {\bf 12}, 199 (1950).

\bibitem{36}  D. L. Hunter and G. A. Baker, Phys. Rev. B {\bf 7}, 3346
(1973).

\bibitem{37}  C. M. Bender and S. Boettcher, J. Math. Phys. {\bf 35}, 1914
(1994).

\bibitem{38}  M. Muthukumar and B. G. Nickel, J. Chem. Phys. {\bf 80}, 5839
(1984).

\bibitem{39}  M. Muthukumar and B. G. Nickel, J. Chem. Phys. {\bf 86}, 460
(1987).

\bibitem{40}  B. Li, N. Madras, and A. D. Sokal, J. Stat. Phys. {\bf 80},
661 (1995).

\bibitem{41}  A. Pelissetto and E. Vicari, cond-mat/0012164 (2000).

\bibitem{42}  D. Sornette, {\it Critical Phenomena in Natural Sciences}
(Springer, Heidelberg, 2000).

\bibitem{43}  A. Helmstetter, D. Sornette, J.-R. Grasso, J. V. Andersen,
S. Gluzman and V. Pisarenko,
Slider-Block Friction Model for Landslides: Implication for Prediction
of Mountain Collapse, submitted to J. Geophysical Research,
cond-mat/0208413.

\bibitem{44}  A. J. Guttman, Phys. Rev. B {\bf 33}, 5089 (1986).

\bibitem{45}  M. N. Barber, R. B. Pearson, D. Toussaint, and J. L.
Richardson, Phys. Rev. B {\bf 32}, 1720 (1985).

\bibitem{46}  H. X. He, C. J. Hamer, and J. Oitmaa, J. Phys. A {\bf 23},
1775 (1990).

\bibitem{47}  S. Gluzman, D. Sornette, and V.I. Yukalov, cond-mat/0204326
(2002).

\bibitem{algocompl}  Chaitin, G.J. {\it Algorithmic Information Theory} 
(Cambridge University, Cambridge, 1987).

\end{thebibliography}
\end{document}